\documentstyle[12pt]{article}

\begin{document}

\setcounter{page}{1}

\pagestyle{plain} \vspace{1cm}

\begin{center}
{\bf A note on Hilbert Space Representation of Quantum Mechanics with Minimal Length}\\
\scriptsize
\vspace{1cm}
{\bf Amir Etemadi}\footnote{amir.etemadi13@gmail.com}\quad  and \quad {\bf Kourosh Nozari}\footnote{knozari@umz.ac.ir}\\
{\it Department of Physics, Faculty of Basic Sciences\\
University of Mazandaran, P. O. Box 47416-95447\\
Babolsar, IRAN}

\end{center}

\vspace{1.5cm}

\begin{abstract}

We study some fundamental issues related to the Hilbert space
representation of quantum mechanics in the presence of a minimal
length and maximal momentum. In this framework, the maximally
localized states and quasi-position representation introduced by
Kempf \emph{et al.} are reconsidered and modified. We show that all
studies in recent years, including \cite{Kempf 1995} and
\cite{Nozari 2012} need serious modification in order to be a
consistent framework for quantum
mechanics in Planck scale.\\
\end{abstract}\vspace{10mm}
\newpage

\section{\large{\textbf{$\!\!\!\!\!\!\!\!$.$\;\;$Introduction}}}

It seems that a natural ultraviolet cutoff, a minimum distance
$\ell_{min}$, is an inevitable prediction of all approaches to
quantum gravity proposal \cite{Veneziano 1986}-\cite{Adler 2010}.
This is due to powerful gravitational effects on the very structure
of spacetime when we aim to resolve very small distances. In the
context of the doubly special relativity it is shown that the
presence of a minimal measurable length will follow the appearance
of a maximum measurable momentum for test particles \cite{Magueijo
2002}-\cite{Cortes 2005}. This is in fact in complete agreement with
the notion of the uncertainty principle. Therefore, in quantum
gravity regime there are a lower bound for position measurements and
consequently an upper bound for momentum measurements. Minimal
length and maximal momentum modify the Heisenberg Uncertainty
Principle (HUP) to the so-called Gravitational/Generalised
Uncertainty Principle (GUP) and therefore, a revision of the
standard Heisenberg algebra is inevitable. A generalized Heisenberg
algebra in the presence of both minimal length and maximal momentum
can be formulated as \cite{Ali 2009}-\cite{Ali 2011}
\begin{equation}
[x_{i},p_{j}] =
i\hbar\Big(\delta_{ij}-\alpha(p\delta_{ij}+\frac{p_{i}p_{j}}{p})+\alpha^2(p^2\delta_{ij}+3p_{i}p_{j})\Big)
\end{equation}
where $\alpha=\frac{\alpha_{0}}{M_{Pl}c}$ is the GUP parameter and
$M_{Pl}\sim10^{-8}$ Kg is the Planck mass. Note that dimensionally
$[\alpha]=(momentum)^{-1}$, and $\alpha_0$ is a dimensionless
quantity. It is normally assumed that $\alpha_0$ is of the order of
unity, $\alpha_0\sim1$. In this case the $\alpha$-dependent terms
are important only for energies near the Planck scale $\sim10^{19}$
GeV. In one dimension, Eq.(1) up to ${\mathcal{O}}(\alpha^2)$ terms
follows the GUP
\begin{eqnarray}
\Delta x\Delta p\geq\frac{\hbar}{2}[1-2\alpha\langle p\rangle+4\alpha^2\langle p^2\rangle]\hspace{44mm}\nonumber\\
\geq\frac{\hbar}{2}\Big[1+\Big(\frac{\alpha}{\sqrt{\langle
p^2\rangle}}+4\alpha^2\Big)({\Delta p})^2+4\alpha^2\langle
p\rangle^2- 2\alpha\sqrt{\langle p^2\rangle}\Big]\,,
\end{eqnarray}
where unlike the ordinary Heisenberg Uncertainty Principle, one can
no longer make $\Delta x$ arbitrarily small by letting $\Delta p$ to
grow arbitrarily. The GUP obtained in Eq.(2) implies a minimum
measurable length and also a maximum measurable momentum as
$$\Delta x_{min}\approx\alpha_0\ell_{Pl}$$
$$\Delta p_{max}\approx\frac{M_{Pl}c}{\alpha_0}$$
where $\ell_{Pl}\sim10^{-35}$ m is the Planck length. Due to the
existence of these two natural cutoffs, several modifications would
be appeared in the very basics of the standard quantum mechanics and
these modifications lead one to a generalized quantum mechanics in
Planck scale. These types of studies are phenomenological in essence
since there is no completely formulated quantum theory of gravity.
Albeit, recently it has been shown in \cite{Nozari 2014} that these
natural cutoffs are actually global (topological) properties of
compact symplectic manifolds much in the same way as gravity is a
global property of curved spacetime. Many formalisms have been
proposed in recent years, but despite all these efforts, yet there
is no complete framework indicating a concrete picture of Planck
scale modified quantum mechanics. One of the most famous approaches
which has been the basis of many researches in recent years, is the
KMM formalism \cite{Kempf 1995}, that is presented by respecting a
GUP with just a minimal measurable length. In this streamline, the
authors of the present paper have generalized the KMM formalism to a
more general case in which the maximal momentum is taken into
account too \cite{Nozari 2012}. But, none of these two frameworks
have been able thoroughly to provide a proper playground of quantum
mechanics in quantum gravity regime. In fact, the main shortcoming
of the mentioned two studies is that they cannot recover the
standard results in the limit of energies much less than the Planck
scale energy, and hence, in the language of the correspondence
principle these studies need to be modified severely. Now, in this
paper, using a generalized Heisenberg algebra defined as
\begin{equation}
[x,p] = i\hbar(1-m\alpha p+n\alpha^2p^2)
\end{equation}
and also taking a maximal momentum as
$P_{max}=\frac{1}{\kappa\alpha}$, we focus on and reconsider the
basic results obtained in the Refs.\cite{Kempf 1995} and
\cite{Nozari 2012}. Then, by a novel analysis we derive a precise
framework for quantum mechanics in extremely high energy regime near
the Planck scale. In this direction, the generalized relations for
plane wave, Dirac $\delta$-function, Fourier transformation, de
Broglie equation and Planck relation would be obtained or modified.
These are, though very simple, some basic and important achievements
leading us to a phenomenologically reliable gravitational quantum
mechanics. These results open also some new windows on the issue of
special relativity in quantum gravity domain (Planck scale).

\section{\large{\textbf{$\!\!\!\!\!\!\!\!$.$\;\;$Minimum Length and Maximal Localization states}}}
As a common feature of all quantum gravity candidate theories, there
is a fundamental length of the order of the Planck length in which
one cannot probe distances smaller than this natural cutoff. This
means that the very notion of localizability should be restricted to
a lower bound (of the order of the Planck length) and there is no
further localization possible in essence. Hence, we are forced to
introduce the maximally localized states \cite{Kempf
1995}-\cite{Kempf 1997} with $\Delta x_{min}=\ell_{min}$ instead of
the usual absolute localized states with $\Delta x_{min}=0$. As has
been mentioned in Ref. \cite{Kempf 1995}, due to the presence of a
nonzero minimum measurable distance, the ordinary position space
representation is no longer applicable in quantum gravity regime.
But, there still exists a continuous momentum space representation
in which we can explore the physical implications of the minimal
length scenario.

We start by defining the operators $\hat{\mathcal P}$ and
$\hat{\mathcal X}$ as (see \cite{Kempf 1995} and \cite{Nozari 2012})
\begin{eqnarray}
\hat{{\mathcal P}}\,\psi(p)=p\,\psi(p)\hspace{38mm}\nonumber\\
\hat{{\mathcal X}}\,\psi(p)=(1-m\alpha p+n{\alpha}^2 p^2)\,\hat{x}\,\psi(p)\hspace{5mm}\nonumber\\
=(1-m\alpha p+n{\alpha}^2 p^2)i\hbar\,\partial_{p}\psi(p)\,
\end{eqnarray}
where $\hat{x}$ and $\hat{p}$ satisfy the canonical commutation
relation $[\hat{x},\hat{p}]=i\hbar$, and $\hat{{\mathcal P}}$ and
$\hat{{\mathcal X}}$ satisfy (3). It is shown that, the scalar
product in this representation should be modified as
\begin{equation}
\langle\Psi|\Phi\rangle=\int_{-P_{max}}^{+P_{max}}\Psi^*(p)\,\Phi(p)\,\frac{dp}{1-m\alpha p+n{\alpha}^2 p^2}\,.
\end{equation}
Further, the generalized identity operator and the generalization of
the scalar product of momentum eigenstates would be represented as
\begin{equation}
\textbf{1}=\int_{-P_{max}}^{+P_{max}}|p\rangle\langle p| \frac{dp}{1-m\alpha p+n\alpha^2 p^2}
\end{equation}
and
\begin{equation}
\langle p|p'\rangle=\big(1-m\alpha p+n\alpha^2 p^2\big)\,\delta(p-p')\,.
\end{equation}
respectively. In order to calculate the states $|\zeta^{ml}\rangle$
of maximal localization around a position $\zeta\geq\ell_{min}$
\begin{equation}
\langle\zeta^{ml}|{\mathcal X}|\zeta^{ml}\rangle = \zeta
\end{equation}
we can use the positivity of the norm \cite{Kempf 1995}
\begin{equation}
\|\Big({\mathcal X}-\langle{\mathcal
X}\rangle+\frac{\langle[{\mathcal X},{\mathcal
P}]\rangle}{2(\Delta{\mathcal P})^2}({\mathcal P} -\langle{\mathcal
P}\rangle)\Big)|\varphi \rangle\|\geq 0\,.
\end{equation}
Considering (3) and $P_{max}=\frac{1}{\kappa\alpha}$, on the
boundary of the physically allowed region, we obtain the states of
maximal localization $\varphi^{ml}_\zeta(p)$ as
\begin{equation}
\varphi^{ml}_\zeta(p)={\mathcal{N}}\,(1-m\alpha p+n\alpha^2p^2)^{-\frac{n+\kappa^2}{4n}}
e^{-\frac{1}{\sqrt{4n-m^2}}\big(\frac{m\alpha(n+\kappa^2)}{2n}+i\frac{2\zeta}{\alpha\hbar}\big)
\big(\tan^{-1}(\frac{2n\alpha p-m}{\sqrt{4n-m^2}})+\tan^{-1}(\frac{m}{\sqrt{4n-m^2}})\big)}\,.
\end{equation}
in which ${\mathcal{N}}$ is the normalization factor (one can find
the details of computations in \cite{Nozari 2012}). Note that, these
states are obtained for $\langle p\rangle=0$ and $\Delta
p=\frac{1}{\kappa\alpha}$ that gives the states of absolutely
maximal localization and critical momentum uncertainty
(corresponding to the maximal momentum) respectively. In the
language of Dirac notation, $\varphi^{ml}_\zeta(p)$ can be written
as $\langle p|\zeta^{ml}\rangle$ which presents the probability
amplitude for the particle with the momenta $p$, being maximally
localized around the ``position'' $\zeta$. Thus
${\varphi^{ml}_\zeta(p)}$ (or
${\varphi^{ml}_\zeta(p)}^{*}=\langle\zeta^{ml}|p\rangle$) gives the
generalized concept of change of basis or the ordinary translation
function $\langle p|x\rangle$ (or $\langle x|p\rangle$).

Now the critical point which has been the basis of mistakes in
previous studies (\cite{Kempf 1995} and \cite{Nozari 2012}) shows
itself: in these studies the authors have used the relation
$\langle\zeta^{ml}|\zeta^{ml}\rangle=1$ for normalization of the
maximally localized states. Unfortunately, this procedure has led
the authors of these papers to a normalization factor that vanishes
in the limit of the standard quantum mechanics. As a result, KMM in
\cite{Kempf 1995} found a divergent energy for a test particle in
the limit of the standard quantum mechanics, which is obviously
impossible! This is more or less in the same manner in Ref.
\cite{Nozari 2012}, though one more step has been taken toward the
complete framework. Here, we focus on this issue and present a
deeper argument on this issue to see the essence of the problem and
its possible solution. For this purpose, we use the
\emph{completeness} of the set of maximally localized eigenbasis
$\{|\zeta^{ml}\rangle\}$ (for proof of completeness, see the
appendix of Ref. \cite{Kempf 1997}). By using the completeness
relation in the left hand side of the generalized relation of the
scalar product of momentum eigenstates (7), we obtain
\begin{equation}
\int_{-\infty}^{+\infty}\varphi^{ml}_\zeta(p)\,{\varphi^{ml}_\zeta}^*(p')\,d\zeta=(1-m\alpha p+n\alpha^2 p^2)\,\delta(p-p')\,.
\end{equation}
Then, from Eq. (10) it follows that
\begin{eqnarray}
{\mathcal{N}}{{\mathcal{N}}}^*\,\frac{e^{-\frac{m\alpha(n+\kappa^2)}{2n\sqrt{4n-m^2}}\big(\tan^{-1}(\frac{2n\alpha p-m}
{\sqrt{4n-m^2}})+\tan^{-1}(\frac{2n\alpha p'-m}{\sqrt{4n-m^2}})+2\tan^{-1}(\frac{m}{\sqrt{4n-m^2}})\big)}}{(1-m\alpha p+
n\alpha^2p^2)^{\frac{n+\kappa^2}{4n}}\,(1-m\alpha p'+n\alpha^2p'^2)^{\frac{n+\kappa^2}{4n}}}\times\hspace{40mm}\nonumber\\
\frac{1}{(1-m\alpha p+n\alpha^2 p^2)}\int_{-\infty}^{+\infty}e^{-i\frac{\zeta}{\hbar}\frac{2}{\alpha\sqrt{4n-m^2}}\big(\tan^{-1}
(\frac{2n\alpha p-m}{\sqrt{4n-m^2}})-\tan^{-1}(\frac{2n\alpha p'-m}{\sqrt{4n-m^2}})\big)}d\zeta=\delta(p-p')\,.
\end{eqnarray}
Taking into account the general property of Dirac $\delta$-function
$\delta(\Omega(z))=\frac{1}{\Omega'(z_0)}\delta(z-z_0)$, it would be
obtained that
\begin{equation}
\int_{-\infty}^{+\infty}e^{-i\frac{\zeta}{\hbar}\frac{2}{\alpha\sqrt{4n-m^2}}
\big(\tan^{-1}(\frac{2n\alpha p-m}{\sqrt{4n-m^2}})-\tan^{-1}(\frac{2n\alpha p'-m}{\sqrt{4n-m^2}})\big)}d\zeta=
2\pi\hbar\,(1-m\alpha p'+n\alpha^2p'^2)\,\delta(p-p')\,.
\end{equation}
By putting this in (12) we obtain the normalization factor as
\begin{equation}
{\mathcal{N}}=\frac{1}{\sqrt{2\pi\hbar}}\,(1-m\alpha p+n\alpha^2p^2)^{\frac{n+\kappa^2}{4n}}\,e^{\frac{m\alpha(n+\kappa^2)}
{2n\sqrt{4n-m^2}}\big(\tan^{-1}(\frac{2n\alpha p-m}{\sqrt{4n-m^2}})\,+\,\tan^{-1}(\frac{m}{\sqrt{4n-m^2}})\big)}\,.
\end{equation}
Therefore, the momentum space wave functions $\varphi^{ml}_\zeta(p)$
which are maximally localized around a position $\zeta$, would be
achieved as
\begin{equation}
\varphi^{ml}_\zeta(p)=\frac{1}{\sqrt{2\pi\hbar}}\,e^{-i\frac{2\zeta}{\alpha\hbar\sqrt{4n-m^2}}
\big(\tan^{-1}(\frac{2n\alpha p-m}{\sqrt{4n-m^2}})\,+\,\tan^{-1}(\frac{m}{\sqrt{4n-m^2}})\big)}\,.
\end{equation}
This is completely different with the results obtained in previous
studies (see \cite{Kempf 1995}, Eq. (37) and \cite{Nozari 2012}, Eq.
(35)). Now, there is a correct limiting result for
$\alpha\rightarrow0$ in the favor of correspondence principle. Note
that the normalization factors in previous studies were vanishing in
this limit which cannot be the case based on the correspondence
principle. Now equation (15) gives the generalized profile of the
plane wave solution as
$$e^{i\frac{2\zeta}{\alpha\hbar\sqrt{4n-m^2}}\big(\tan^{-1}(\frac{2n\alpha p-m}{\sqrt{4n-m^2}})\,+\,\tan^{-1}(\frac{m}{\sqrt{4n-m^2}})\big)}\,.$$
One can easily check that, in the limit of $\alpha\rightarrow0$ the
ordinary plane wave profile and momentum space wave function would
be exactly recovered
\begin{equation}
\lim_{\alpha\rightarrow 0}\
\frac{1}{\sqrt{2\pi\hbar}}\,e^{-i\frac{2\zeta}{\alpha\hbar\sqrt{4n-m^2}}\big(\tan^{-1}(\frac{2n\alpha
p-m}{\sqrt{4n-m^2}})\,+
\,\tan^{-1}(\frac{m}{\sqrt{4n-m^2}})\big)}\Rightarrow
\frac{1}{\sqrt{2\pi\hbar}}\,e^{-i\frac{px}{\hbar}}
\end{equation}
or
\begin{equation}
\lim_{\alpha\rightarrow 0}\ \langle
p|\zeta^{ml}\rangle\Rightarrow\langle p|x\rangle\,.
\end{equation}

In comparison with ordinary wave mechanics, the generalized
relations of the plane wave and momentum space wave function lead us
to a significant outcome; the modified wavenumber in quantum gravity
regime ${\mathcal{K}}_{QG}$ as
\begin{equation}
{\mathcal{K}}_{QG}=\frac{2}{\alpha\hbar\sqrt{4n-m^2}}\Big(\tan^{-1}
(\frac{2n\alpha
p-m}{\sqrt{4n-m^2}})\,+\,\tan^{-1}(\frac{m}{\sqrt{4n-m^2}})\Big)\,.
\end{equation}
Therefore, the modified form of the corresponding wavelength would
be resulted as
\begin{equation}
\lambda_{QG}=\frac{\pi\alpha\hbar\sqrt{4n-m^2}}{\tan^{-1}(\frac{2n\alpha p-m}
{\sqrt{4n-m^2}})\,+\,\tan^{-1}(\frac{m}{\sqrt{4n-m^2}})}\,.
\end{equation}
So, for massless particles we can infer a generalized frequency as
follows
\begin{equation}
\nu_{QG}=\frac{c}{\pi\alpha\hbar\sqrt{4n-m^2}}\Big(\tan^{-1}(\frac{2n\alpha
p-m} {\sqrt{4n-m^2}})\,+\,\tan^{-1}(\frac{m}{\sqrt{4n-m^2}})\Big)\,.
\end{equation}
Now, by keeping $h$ as a subatomic-scale constant which describes
the relationship between energy and frequency as
$$\frac{\textrm{Energy}}{\textrm{Frequency}}=h$$
we arrive at the generalized Planck relation in the domain of
quantum gravity as ${\mathcal{E}}_{QG}=h\,\nu_{QG}$. The presence of
a maximum measurable momentum concludes that there is no wavelength
smaller than $\lambda_{QG}(P_{max})$, or equivalently no frequency
larger than $\nu_{QG}(P_{max})$. Hence, the highest energy for a
massless particle would be
\begin{eqnarray}
{\mathcal{E}}_{QG}(P_{max})=h\,\nu_{QG}(P_{max})\hspace{36mm}\nonumber\\
=\frac{2c}{\alpha\sqrt{4n-m^2}}\tan^{-1}\Big(\frac{\sqrt{4n-m^2}}{2\kappa-m}\Big)\,\,.
\end{eqnarray}
Further, Eqs. (18) and (19) together with the generalized Planck
relation lead us to the ``generalized de Broglie relation''
${\mathcal{P}}_{QG}$ as
\begin{equation}
{\mathcal{P}}_{QG}=\frac{2}{\alpha\sqrt{4n-m^2}}\Big(\tan^{-1}(\frac{2n\alpha
p-m}{\sqrt{4n-m^2}})\,+\,\tan^{-1}(\frac{m}{\sqrt{4n-m^2}})\Big)\,.
\end{equation}
We call this quantity as ``\textit{quasimomentum}\," in what
follows, since it has the dimension of momentum. The results
obtained so far contain a crucial point that in extremely high
energy regime the role of momentum should be reconsidered
essentially in comparison with the standard situation. Indeed, the
quantity quasimomentum does not mean just a modified momenta here.
It can be interpreted as the modified method of the momentum
arrangement in the related equations. As a result, one encounters
${\mathcal{P}}_{QG}(p)$ instead of $p$ in field equations.
Accordingly, one concludes the modified kinetic energy for particles
as follows
\begin{eqnarray}
{\mathcal{E}}^{kin}_{QG}(p)=\frac{[{\mathcal{P}}_{QG}(p)]^2}{2M}\hspace{91mm}\nonumber\\
=\frac{2}{\alpha^2M(4n-m^2)}\Big(\tan^{-1}(\frac{2n\alpha
p-m}{\sqrt{4n-m^2}})\,+\,\tan^{-1}(\frac{m}{\sqrt{4n-m^2}})\Big)^2\,.
\end{eqnarray}
This is a new approach and completely different from the
considerations adopted in previously proposed formalisms. This
result leads us to a new formulation of energy in quantum gravity
regime. Since there is an upper bound for momentum, $P_{max}$, so
the most energetic particles would have the kinetic energy as
${\mathcal{E}}^{kin}_{QG}(P_{max})$. It is easy to check that all
these relations in the limit of $\alpha\rightarrow0$ recover the
corresponding ordinary relations.

For the expectation value of energy ${\mathcal{E}}^{kin}_{QG}$ in
this setup we have
\begin{eqnarray}
\langle\zeta^{ml}|\frac{[{\mathcal{P}}_{QG}(p)]^2}{2M}|\zeta^{ml}\rangle=\frac{2}{\alpha^2M(4n-m^2)}
\int_{-P_{max}}^{+P_{max}}\frac{\Big(\tan^{-1}(\frac{2n\alpha
p-m}{\sqrt{4n-m^2}})\,+
\,\tan^{-1}(\frac{m}{\sqrt{4n-m^2}})\Big)^2}{1-m\alpha p+n\alpha^2p^2}\,dp\nonumber\\
=\frac{8}{3\alpha^3M(4n-m^2)^{\frac{3}{2}}}\Big(\tan^{-1}(\frac{\sqrt{4n-m^2}}{2\kappa-M})\Big)^3\hspace{43mm}
\end{eqnarray}
which indicates that in contrast to ordinary states, the maximal
localization states are proper physical states with finite energy.
We note that this relation in the limit of $\alpha\rightarrow0$ goes
to infinity and this is not surprising since now there is no
restriction on momentum values in the same way as the standard
quantum mechanics. The scalar product of the maximally localized
states now is given by
\begin{equation}
\langle\zeta^{ml}|{\zeta'\,}^{ml}\rangle=\frac{1}{2\pi\hbar}\int_{-P_{max}}^{+P_{max}}
e^{i\frac{2(\zeta-\zeta')}{\alpha\hbar\sqrt{4n-m^2}}\big(\tan^{-1}(\frac{2n\alpha p-m}{\sqrt{4n-m^2}})\,+
\,\tan^{-1}(\frac{m}{\sqrt{4n-m^2}})\big)}\frac{dp}{1-m\alpha p+n\alpha^2p^2}\,.
\end{equation}
Here, we define the generalized Dirac $\delta$-function in the
modified quantum mechanics as
\begin{equation}
\delta(\zeta-\zeta')=\frac{1}{2\pi}\int_{-P_{max}}^{+P_{max}}e^{i{\mathcal{K}}_{QG}(\zeta-\zeta')}\,d{\mathcal{K}}_{QG}
\end{equation}
whence we obtain
\begin{equation}
\langle\zeta^{ml}|{\zeta'\,}^{ml}\rangle=\delta(\zeta-\zeta')\,.
\end{equation}
So, unlike the previous studies (\cite{Kempf 1995} and \cite{Nozari
2012}), now there is mutual orthogonality of the maximal
localization states! Indeed, the maximal localization state
$|\zeta^{ml}\rangle$ and its momentum space counterpart
$\varphi^{ml}_\zeta(p)$, provide a proper background for describing
the behavior of particles near the Planck scale. In a similar
fashion, we need to change and modify our viewpoint on the very
notion of space too. That is to say, we need a modified position
space which realizes the existence of a minimum distance in its very
structure from the beginning. Actually, in order to work with the
maximal localization states, one needs a generalized space that
treats the minimal length as an ultimate limit for the resolution of
the spacetime points or nearby particles.

\section{\large{\textbf{$\!\!\!\!\!\!\!\!$.$\;\;$Quasiposition Space}}}
In ordinary quantum mechanics one has the position and momentum
space representations in Hilbert space with position and momentum
wave functions given as $\psi(x)=\langle x|\psi\rangle$ and
$\psi(p)=\langle p|\psi\rangle$ respectively. But, in extreme
situations such as the Planck scale, this framework would be
drastically disturbed because of the presence of a nonzero minimum
measurable length. When there exists a minimal length, it means that
there is a nonzero uncertainty in position measurements as
\begin{equation}
(\Delta x)^2 _{|\psi>}=\big<\psi|(X-\big<\psi|X|\psi\big>)^2|\psi\big>\geq\Delta x_{min}\,.
\end{equation}
So, in one hand, we should change the ordinary concept of absolute
localizability $\Delta x_{min}=0$ to the modified concept of maximal
localization, i.e. the states that are localized just up to the
minimal length $\ell_{min}$. On the other hand, we can no longer
build a Hilbert space on the usual position wave function, and thus,
the ordinary position space has no sense in this respect
\cite{Nozari 2012}-\cite{Kempf 1994}. Hence, in extremely high
energy regimes we need to reformulate quantum mechanics in a
generalized space with minimal length. In order to work with
maximally localized states $|\zeta^{ml}\rangle$, we need a space in
which the concept of point or localizability is modified in the
presence of a minimal length. In this sense, the quasiposition space
introduced by KMM formalism \cite{Kempf 1995} would be the proper
representation. In fact, quasiposition space is the modified notion
of the ordinary position space which treats the existence of a
minimal length in a realistic manner.

Taking $|\phi\rangle$ as an arbitrary state, we can define
$\langle\zeta^{ml}|\phi\rangle$ as the state's quasiposition wave
function $\phi(\zeta)$ \cite{Kempf 1995}, \cite{Nozari 2012}. That
is, $\phi(\zeta)$ projects the probability amplitude for the
particle being maximally localized around the position $\zeta$ in
the quasiposition space. So, from Eqs. (6) and (15), the
transformation for a state wave function in the momentum
representation, $\phi(p)=\langle p|\phi\rangle$, to its
quasiposition wave function is
\begin{eqnarray}
\phi(\zeta)=\frac{1}{\sqrt{2\pi\hbar}}\int_{-P_{max}}^{+P_{max}}e^{i\frac{2\zeta}
{\alpha\hbar\sqrt{4n-m^2}}\big(\tan^{-1}(\frac{2n\alpha p-m}{\sqrt{4n-m^2}})\,+
\,\tan^{-1}(\frac{m}{\sqrt{4n-m^2}})\big)} \phi(p) \frac{dp}{1-m\alpha p+n\alpha^2p^2}\,.
\end{eqnarray}
This transformation explicitly exhibits the generalization of the
Fourier transformation. By inverse Fourier transform, we have the
transformation of a quasiposition wave function into a momentum
space wave function as
\begin{eqnarray}
\phi(p)=\frac{1}{\sqrt{2\pi\hbar}}\int_{-\infty}^{+\infty}e^{-i\frac{2\zeta}{\alpha\hbar\sqrt{4n-m^2}}
\big(\tan^{-1}(\frac{2n\alpha p-m}{\sqrt{4n-m^2}})\,+\,\tan^{-1}(\frac{m}{\sqrt{4n-m^2}})\big)} \phi(\zeta)\,d\zeta\,.
\end{eqnarray}
The remarkable note is that, here unlike the prior formalisms, in
the limit of $\alpha\rightarrow0$ we exactly recover the
corresponding ordinary transformations (this is not the case for KMM
framework for instance). Now, using Eqs. (5) and (30), we can
calculate the scalar product of two arbitrary states $|\phi\rangle$
and $|\psi\rangle$ in terms of the quasiposition wave functions
$\phi(\zeta)$ and $\psi(\zeta)$ as
\begin{eqnarray}
\langle\phi|\psi\rangle=\int_{-P_{max}}^{+P_{max}}\phi^*(p)\psi(p) \frac{dp}{1-m\alpha p+n\alpha^2p^2}\hspace{48mm}\nonumber\\
=\frac{1}{2\pi\hbar}\int_{-P_{max}}^{+P_{max}}\!\!\!\int_{-\infty}^{+\infty}\!\!\!
\int_{-\infty}^{+\infty}e^{i\frac{2(\zeta-\zeta')} {\alpha\hbar\sqrt{4n-m^2}}\big(\tan^{-1}
 (\frac{2n\alpha p-m}{\sqrt{4n-m^2}})+\tan^{-1}(\frac{m}{\sqrt{4n-m^2}})\big)}\phi^*(\zeta)
 \psi(\zeta')d\zeta d\zeta'\frac{dp}{1-m\alpha p+n\alpha^2p^2}\nonumber\\
=\int_{-\infty}^{+\infty}\phi^*(\zeta)\psi(\zeta)d\zeta\hspace{80mm}
\end{eqnarray}
where we have used the generalized Dirac $\delta$-function as given
by (26).

From Eq. (4) we have the operators $\hat{{\mathcal
P}}=\frac{\hbar}{i}\partial_{\zeta}$ and $\hat{{\mathcal
X}}=(1-m\alpha p+n{\alpha}^2 p^2)i\hbar\,\partial_{p}$. By applying
these operators on the generalized plane wave (15) we can derive
\begin{eqnarray}
\hat{{\mathcal P}}\,\varphi^{ml}_\zeta(p)={\mathcal P}_{QG}\,\varphi^{ml}_\zeta(p)\nonumber\\
\hat{{\mathcal X}}\,\varphi^{ml}_\zeta(p)=\zeta\,\varphi^{ml}_\zeta(p)\hspace{5mm}\,
\end{eqnarray}
Since $\partial_{p}\equiv\frac{1}{(1-m\alpha p+n{\alpha}^2 p^2)}
\partial_{{\mathcal{P}}_{QG}}$\,, so we can represent operator $\hat{{\mathcal X}}$ as
$i\hbar\partial_{{\mathcal{P}}_{QG}}$ which can be called as
"\textit{quasiposition operator}". Therefore, in this generalized
framework momentum and quasiposition operators operate as
\begin{eqnarray}
\hat{{\mathcal P}}\,\phi(\zeta)=\frac{\hbar}{i}\partial_{\zeta}\,\phi(\zeta)\nonumber\\
\hat{{\mathcal X}}\,\phi(\zeta)=\zeta\,\phi(\zeta)\hspace{5mm}
\end{eqnarray}
in quasiposition representation, and also as
\begin{eqnarray}
\hat{{\mathcal P}}\,\phi(p)={\mathcal P}_{QG}\,\phi(p)\hspace{5mm}\nonumber\\
\hat{{\mathcal X}}\,\phi(p)=i\hbar\,\partial_{{\mathcal{P}}_{QG}}\,\phi(p)
\end{eqnarray}
in momentum space representation. These are novel achievement in
comparison with the corresponding results obtained in \cite{Kempf
1995} and also \cite{Nozari 2012}. Here we obtained the appropriate
operators $\hat{{\mathcal P}}$ and $\hat{{\mathcal X}}$ which act
straightforwardly on the wave functions in momentum and
quasiposition representations. As it is obtained in Ref.
\cite{Nozari 2012}, we see that here there is not noncommutativity
in the structure of quasiposition space $[{\mathcal{X}}_i,
{\mathcal{X}}_j]=0$. Therefore, quasiposition space now is a proper
space in order to study the maximal localization states in Planck
scale quantum mechanics.

\section{\large{\textbf{$\!\!\!\!\!\!\!\!$.$\;\;$Implications for Special Relativity}}}

Now by having the results obtained in previous sections in hand, we
focus on the possible implications of these results on special
relativity. In special relativity one has
$$E=\sqrt{p^2c^2+E_0^2}\,.$$ For massless particles, $M_0=0$, and therefore $E=pc$.
In our case the generalized de Broglie relation leads to the
following relation for the generalized relativistic energy of
photons and other massless particles
\begin{eqnarray}
{\mathcal{E}}^{Rel}_{QG}={\mathcal{P}}_{QG}\,c\hspace{89mm}\nonumber\\
=\frac{2c}{\alpha\sqrt{4n-m^2}}\Big(\tan^{-1}(\frac{2n\alpha
p-m}{\sqrt{4n-m^2}})\,+\,\tan^{-1}(\frac{m}{\sqrt{4n-m^2}})\Big)\,,
\end{eqnarray}
where recovers the standard relation $E=pc$ in the limit of
$\alpha\rightarrow0$. For massive particles we have
\begin{eqnarray}
{\mathcal{E}}^{Rel}_{QG}=\sqrt{{\mathcal{P}}_{QG}^2c^2+E_0^2}\hspace{91mm}\nonumber\\
=\sqrt{\frac{4c^2}{\alpha^2(4n-m^2)}\big(\tan^{-1}(\frac{2n\alpha
p-m}
{\sqrt{4n-m^2}})\,+\,\tan^{-1}(\frac{m}{\sqrt{4n-m^2}})\big)^2+E_0^2}\,,
\end{eqnarray}
where $$\lim_{\alpha\rightarrow
0}{\mathcal{E}}^{Rel}_{QG}=\sqrt{p^2c^2+E_0^2}$$ in the favor of the
correspondence principle. To proceed further, we look at the
modification of the Lorentz factor
$\gamma(v)=\frac{1}{\sqrt{1-(\frac{v}{c})^2}}$ in our framework.
From the relativistic momentum equation
$p(v)=\frac{M_0v}{\sqrt{1-(\frac{v}{c})^2}}$, one can obtain easily
\begin{equation}
\gamma(p)={\sqrt{1+(\frac{p}{M_0c})^2}}.
\end{equation}
Then, putting the quasimomentum in this relation we deduce the
generalized Lorentz factor in our framework as follows
\begin{equation}
\gamma_{QG}(p)={\sqrt{1+\frac{4}{M_0^2c^2\alpha^2(4n-m^2)}\Big(\tan^{-1}(\frac{2n\alpha
p-m}
{\sqrt{4n-m^2}})\,+\,\tan^{-1}(\frac{m}{\sqrt{4n-m^2}})\Big)^2}}.
\end{equation}
One could reach this generalized form of the Lorentz factor from the
energy-momentum relation $E=\gamma(p)E_0$ too. As the most essential
factor in all relativistic formulae, this generalized factor would
modify the basic relations of special relativity. Thus, in extremely
high energy regimes one has generalized relativistic equations such
as the time dilation as ${\mathcal{T}}_{QG}=\gamma_{QG}(p)\,T_0$ and
length contraction as
${\mathcal{L}}_{QG}=\frac{L_0}{\gamma_{QG}(p)}$. While in the
standard special relativity for velocities near the speed of light
one has
\begin{equation}
\lim_{v\rightarrow
c}\gamma(v)=\lim_{p\rightarrow\infty}\gamma(p)=\infty\,,
\end{equation}
in the generalized framework presented here, due to the presence of
a maximal momentum, one has
\begin{equation}
\lim_{p\rightarrow
P_{max}}\gamma_{QG}(p)={\sqrt{1+\frac{4}{M_0^2c^2\alpha^2(4n-m^2)}
\Big(\tan^{-1}(\frac{\sqrt{4n-m^2}}{2\kappa-M})\Big)^2}}\,.
\end{equation}
In the energies much less than the Planck scale,
$\alpha\rightarrow0$, the ordinary special relativity would be
exactly recovered.

\section{\large{\textbf{$\!\!\!\!\!\!\!\!$.$\;\;$Summary}}}

In this note we have shown that there is a serious flaw in the
renowned paper \cite{Kempf 1995} and then we have provided a
strategy to overcome this flaw. In this framework the generalized
relations for plane wave profile, Dirac $\delta$-function, Fourier
transformation, the de Broglie equation and the Planck relation are
obtained or modified. We have also derived some new and important
relations for special relativity in quantum gravity domain.

\end{document}